\documentstyle[epsfig]{aipproc}

\def\bge{\begin{equation}}
\def\ene{\end{equation}}
\def\bg{\begin{eqnarray}}
\def\en{\end{eqnarray}}
\def\nn{\nonumber} 
\def\vr{\vec{r}}

\begin{document}
\title{Isospin Symmetry Breaking in Nuclei \\ --- ONS Anomaly ---}

\author{Koichi Saito}
\address{Tohoku College of Pharmacy, Sendai 981-8558, Japan}

\maketitle

\begin{abstract}
We study the binding energy differences of the valence proton and  
neutron of the mirror nuclei, $^{15}$O - $^{15}$N, $^{17}$F - $^{17}$O, 
$^{39}$Ca - $^{39}$K and $^{41}$Sc - $^{41}$Ca, 
using the quark-meson coupling model.  
The calculation involves nuclear structure 
and shell effects explicitly. It is shown that 
binding energy differences of a few hundred keV arise from the strong
interaction, even after subtracting all electromagnetic corrections. 
The origin of these differences may be ascribed to the charge symmetry 
breaking effects set in the strong interaction through 
the u and d current quark mass difference.  In this report, 
we first review the quark-meson coupling model.  In particular, 
we discuss about the nucleon mass in nuclear medium.  
Then, we present details of the charge symmetry 
breaking in finite nuclei, especially the Okamoto-Nolen-Schiffer 
anomaly.  
\end{abstract}

\section*{INTRODUCTION}

The discrepancy between the calculated binding energy differences
of mirror nuclei and those measured is a long-standing problem
in nuclear physics. It is known as the
Okamoto-Nolen-Schiffer (ONS) anomaly \cite{oka,nol}. 
Although it was first thought that electromagnetic effects 
could almost account for the observed binding energy differences, 
it is now believed that the ONS anomaly has its origin in 
charge symmetry breaking (CSB) in the strong interaction \cite{mil}. 
In addition to calculations based on charge symmetry violating 
meson exchange potentials \cite{mil,blu,sha}, 
a number of quark-based calculations have been performed \cite{hen,sai}  
in an attempt to resolve this anomaly. In such calculations, CSB enters 
through the up (u) and down (d) current quark mass difference in QCD. 
Despite these efforts, the difficulty of producing a realistic description 
of nuclear structure on the basis of explicit quark degrees of freedom has 
hindered the direct calculation of the binding energy differences. 

In this study we report the results for the binding energy differences 
of the valence (excess) proton and neutron of the mirror nuclei, 
$^{15}$O -- $^{15}$N, $^{17}$F -- $^{17}$O, $^{39}$Ca -- $^{39}$K and 
$^{41}$Sc -- $^{41}$Ca, calculated using a quark-based model 
involving explicit nuclear structure and shell 
effects, namely the quark-meson coupling (QMC) model \cite{gui}. 
This model has been successfully applied not only to 
traditional nuclear problems \cite{gui}   
but also to other new areas as well \cite{tsu}. 
Although some exploratory QMC results on the ONS anomaly 
have already been reported \cite{sai}, an early version of the model was 
used there, and it was applied to finite nuclei only through local 
density approximation, rather than a consistent shell model calculation. 

\section*{THE QUARK-MESON COUPLING MODEL}

In this section, 
we introduce the QMC model, and then report the medium 
modification of the nucleon properties in finite nuclei \cite{gui}.   

\subsection*{Effect of Nucleon Structure}

Let us suppose that a free nucleon (at the origin) consists of three light 
(u and d) quarks under a (Lorentz scalar) confinement potential, $V_c$.  
Then, the Dirac equation for the quark field $\psi_q$ is given by 
\bge
[ i\gamma\cdot\partial - m_q - V_c(r) ] \psi_q(r) = 0 , 
\label{dirac1}
\ene
where $m_q$ is the bare quark mass.  

Next we consider how Equation (\ref{dirac1}) is modified when the 
nucleon is bound 
in static, uniformly distributed (iso-symmetric) nuclear matter.  
In the QMC model \cite{gui} it is assumed that 
each quark feels scalar, $V_s^q$, and vector, $V_v^q$, potentials, which are 
generated by the surrounding nucleons, as well as the confinement potential.  
This assumption seems appropriate when the nuclear density $\rho_B$ is 
near around normal nuclear matter density ($\rho_0 = 0.15$ fm$^{-3}$).  
If we use the mean-field approximation (MFA) for the meson fields, 
Equation (\ref{dirac1}) may be rewritten as 
\bge
[ i\gamma\cdot\partial - (m_q - V_s^q) - V_c(r) 
 - \gamma_0 V_v^q ] \psi_q(r) = 0 . 
\label{dirac2}
\ene
The potentials generated by the medium are constants because the matter 
distributes uniformly. As the nucleon is static, the time-derivative 
operator in the Dirac equation can be 
replaced by the quark energy, $-i \epsilon_q$.  
By analogy with the procedure applied to the nucleon
in QHD \cite{ser}, if we introduce the 
effective quark mass by $m_q^{\star} = m_q - V_s^q$, the Dirac equation 
(\ref{dirac2}) can be rewritten in the same form as that in free space 
with the mass $m_q^{\star}$ and the energy $\epsilon_q - V_v^q$, instead of 
$m_q$ and $\epsilon_q$. 
In other words, the vector interaction has {\em no effect 
on the nucleon structure} except for an overall phase in the quark wave 
function, which gives a shift in the nucleon energy.  This fact 
{\em does not\/} depend on how to choose the confinement potential, $V_c$.  
Then, the nucleon energy at rest in the medium is given by 
$E_N = M_N^{\star}(V_s^q) + 3V_v^q$, 
where the effective nucleon mass $M_N^{\star}$ depends on {\em only the 
scalar potential}.  

We can extend this idea to finite nuclei \cite{gui}.  
Let us suppose that the scalar and vector potentials in 
Equation (\ref{dirac2}) are 
mediated by the $\sigma$ and $\omega$ mesons, and introduce their 
mean-field values, which now depend on position ${\vec r}$, 
by $V_s^q({\vec r}) = g_{\sigma}^q \sigma({\vec r})$ and 
$V_v^q({\vec r}) = g_{\omega}^q \omega({\vec r})$, respectively, where 
$g_{\sigma}^q$ ($g_{\omega}^q$) is the coupling constant of the quark-$\sigma$ 
($\omega$) meson.  Furthermore, we shall add the isovector vector meson,  
$\rho$, and the Coulomb field, $A$, to describe finite nuclei 
realistically.  
Then, the effective Lagrangian density for finite nuclei 
would be given by \cite{gui}  
\bg
{\cal L}_{QMC}&=& \overline{\psi} [i \gamma \cdot \partial 
- M_N^{\star} - g_\omega \omega \gamma_0 
- g_\rho \frac{\tau^N_3}{2} b \gamma_0 
- \frac{e}{2} (1+\tau^N_3) A \gamma_0 ] \psi  \label{qmc-1} \\
&-& \frac{1}{2}[ (\nabla \sigma)^2 + 
m_{\sigma}^2 \sigma^2 ] 
+ \frac{1}{2}[ (\nabla \omega)^2 + m_{\omega}^2 \omega^2 ] 
+ \frac{1}{2}[ (\nabla b)^2 + m_{\rho}^2 b^2 ] 
+ \frac{1}{2} (\nabla A)^2 ,  \nn
\en
where $\psi$ and $b$ are respectively the nucleon and the $\rho$ fields. 
$m_\sigma$, $m_\omega$ and $m_{\rho}$ are respectively 
the masses of the $\sigma$, $\omega$ and $\rho$ mesons. 
$g_\omega$ and $g_{\rho}$ are respectively the $\omega$-N and $\rho$-N 
coupling constants, which are given by 
$g_\omega = 3 g_\omega^q$ and $g_\rho = g_\rho^q$ (where $g_\rho^q$ is 
the quark-$\rho$ coupling constant).  

If we define the field-dependent $\sigma$-N coupling 
constant, $g_\sigma(\sigma)$, by \cite{gui} 
\bge
M_N^{\star}(\sigma({\vec r})) \equiv M_N - g_\sigma(\sigma({\vec r})) 
\sigma({\vec r}) , \label{coup}
\ene
where $M_N$ is the free nucleon mass, it is easy to compare with 
QHD \cite{ser}.  The difference between QMC   
and QHD lies only in the coupling constant $g_\sigma$, which
depends on the scalar field in QMC while it is constant in QHD.  
However, this difference leads to a lot of favorable results \cite{gui}.  

Now we need a model for the structure of the nucleon involved to 
perform actual calculations.  We here use the MIT bag model.  
In the present model, the bag constant, $B$, and the $z$ 
parameter for the nucleon 
are fixed to reproduce the free nucleon mass ($M_N$ = 939 MeV) and 
the free bag radius $R_N$ = 0.8 fm. 
In the following we choose $m_q$ = 5 MeV and set
$m_\sigma$ = 550 MeV, $m_{\omega}$ = 783 MeV and $m_{\rho}$ = 770 MeV. 
(Variations of the quark mass and $R_N$ only lead to 
numerically small changes in the calculated results \cite{gui}.)  
We then find that $B^{1/4}$ = 170.0 MeV and $z$ = 3.295. 

\begin{figure}[b!] 
\centerline{\epsfig{file=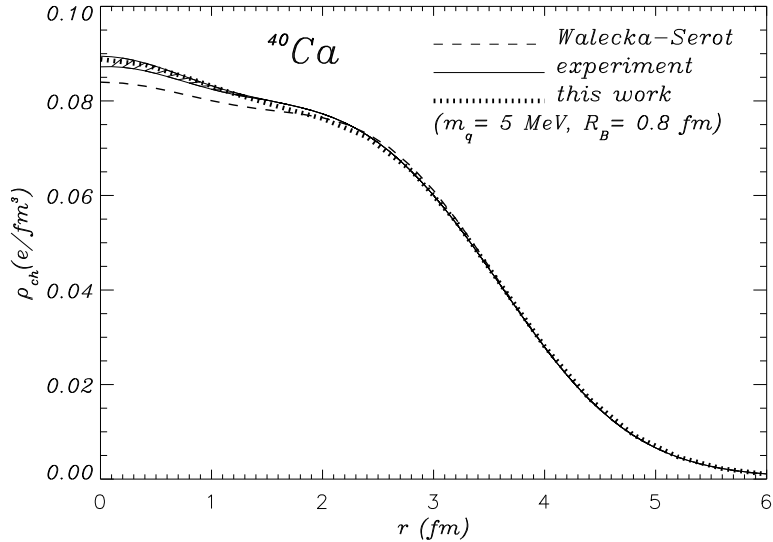,width=11cm}}
\vspace{10pt}
\caption{Charge density distribution for $^{40}$Ca \protect\cite{gui}
compared with the experimental data and that of QHD.}
\label{fig:40Ca}
\end{figure}

For infinite nuclear matter, from the Lagrangian density
(\ref{qmc-1}), we can easily find the total energy per nucleon,  
$E_{tot}/A$, 
and the mean-field values of $\omega$ and $\rho$ (which are respectively 
given by baryon number conservation and the difference 
in proton and neutron densities).  
The scalar mean-field is given by a self-consistency
condition, $\partial E_{tot}/\partial \sigma =0$ \cite{gui}.  
The coupling constants, $g_{\sigma}^2$ and 
$g_{\omega}^2$, are fixed to fit the average 
binding energy ($-15.7$ MeV) at $\rho_0$ for 
nuclear matter.  Furthermore, the $\rho$-N coupling constant is 
used to reproduce the bulk symmetry energy, 35 MeV. 
We then find \cite{gui}:  $g_{\sigma}^2/4\pi$ = 5.40, $g_{\omega}^2/4\pi$
= 5.31, $g_\rho^2/4\pi$ = 6.93, and the nuclear 
incompressibility, $K \simeq 280$ MeV.   
Note that the model gives the variation of the nucleon bag radius, 
$\delta R_N^{\star}/R_N = -0.02$, the lowest quark eigenvalue, 
$\delta x_q^{\star}/x_q = -0.16$ and the root-mean-square radius of the 
quark wave function, $\delta r_q^{\star}/r_q = +0.02$, at saturation density.  

Using these parameters, we can solve a finite nuclear system.  
As an example, we show charge density distribution 
of $^{40}$Ca in Figure \ref{fig:40Ca}.  
The QMC model can reproduce the properties of not only nuclear matter 
but also finite nuclei (for more details, see \cite{gui}).  

\subsection*{Nucleon Mass in Nuclear Matter}

Here we consider the nucleon mass in matter furthermore.  
The nucleon mass is a 
function of the scalar field.  Because the scalar field is small 
at low density the mass may be expanded in terms of $\sigma$ as 
\bge
M_N^{\star} = M_N + \left( \frac{\partial M_N^{\star}}{\partial \sigma} 
\right)_{\sigma=0} \sigma + \frac{1}{2} \left( \frac{\partial^2 M_N^{\star}}
{\partial \sigma^2} \right)_{\sigma=0} \sigma^2 +  {\cal O}({\sigma}^3) . 
\label{nuclm}
\ene
Since the interaction Hamiltonian between the nucleon and the 
$\sigma$ field at the quark level is given by $H_{int} = - 3 g_{\sigma}^q 
\int d{\vec r} \ \overline{\psi}_q \sigma \psi_q$, the derivative of 
$M_N^{\star}$ with respect to $\sigma$ is 
$-3g_{\sigma}^q \int d{\vec r} \ {\overline \psi}_q \psi_q 
\equiv -3g_{\sigma}^q S_N(\sigma)$,  
where we have defined the quark scalar charge in the nucleon, 
$S_N(\sigma)$, which is itself a function of $\sigma$.  
Because of a negative value of the derivative, 
the nucleon mass decreases in matter at low density.  

Furthermore, we define the scalar-charge ratio, $S_N(\sigma)/S_N(0)$,  
to be $C_N(\sigma)$ and the $\sigma$-N coupling constant in free space    
to be $g_\sigma$ (i.e., $g_\sigma = g_\sigma(\sigma=0) 
= 3g_{\sigma}^q S_N(0)$).  Using these quantities, we find  
\bge
M_N^{\star} = M_N - g_{\sigma} \sigma - \frac{1}{2} g_{\sigma} 
C_N^\prime(0) \sigma^2 +  {\cal O}({\sigma}^3) . 
\label{nuclm2}
\ene
In general, $C_N$ is a decreasing function because the quark in matter becomes 
more relativistic than in free space.  Thus, $C_N^\prime(0)$ takes a 
negative value. If the nucleon were structureless $C_N$ would not depend on 
$\sigma$.  Therefore, 
only the first two terms in the RHS of Equation (\ref{nuclm2}) remain, 
which is exactly the same as the effective nucleon 
mass in QHD \cite{ser}.  

\section*{CHARGE SYMMETRY BREAKING IN QMC}

Now we introduce the charge symmetry breaking in 
the QMC model \cite{sai,tsu2}.  
The charge symmetry is explicitly broken at the 
quark level through their masses. 
We use different values for the u and d current quark masses, 
and the effective proton, $M_p^{\star}$, and neutron, $M_n^{\star}$, 
masses.  At position $\vr$ in a 
nucleus (the coordinate origin is taken at the center of the nucleus), 
the Dirac equations for the quarks in the proton or 
neutron bag are given by  
\bge
\left[ i \gamma \cdot \partial_x - \left(
\left(\begin{array}{c} m_u\\ m_d\\ \end{array}\right)
 - V^q_\sigma(\vr) \right) 
- \gamma^0
\left( V^q_\omega(\vr) \pm \frac{1}{2} V^q_\rho(\vr) \right) \right]
\left(\begin{array}{c} \psi_u(x)\\ \psi_d(x)\\ \end{array}\right) = 0, 
\label{diraceq}
\ene
where $|{\vec x} - \vr| \le R_j^{\star}$ ($j$ specifies proton or 
neutron).  
Note that we have assumed that the scalar potential is common to both 
the u and d quarks.  The nucleon and meson fields are  
calculated self-consistently by solving a set of coupled non-linear 
differential equations, derived from the effective Lagrangian density 
(\ref{qmc-1}) with the proper modifications caused by the different proton  
and neutron (or u and d quark) masses in MFA.   
Thus, the present calculation is free from the sort of double counting 
questioned by Auerbach \cite{aue}, and includes the shell effects, which were 
discussed by Cohen et al. \cite{coh}. 

\begin{table}
\caption{Inputs, parameters and some of the quantities calculated in 
the present study. The quantities with a star, $^{\star}$, are those 
quantities calculated at $\rho_0$.  We take $m_u = 5$ MeV.}
\label{parameters}
\begin{tabular}{c|cccccc}
 &$M_j$ (MeV) &$R_j$ (fm) &$B^{1/4}$ (MeV) &$z$ &$M_j^{\star}$ (MeV)
&$R_j^{\star}$ (fm)\\
\tableline 
p (CSB)  &937.6423 (input)&0.8 (input)&169.81&3.305&751.928&0.7950 \\
n (CSB)  &939.6956 (input)&0.8000     &169.81&3.305&753.597&0.7951 \\
N (SU(2))&939.0 (input)   &0.8 (input)&169.97&3.295&754.542&0.7864 \\
\end{tabular}
%
\end{table}

Before discussing the results obtained, we again need to specify the  
parameters and inputs used in the calculation \cite{tsu2}.  
They are summarized in Table \ref{parameters}. 
The bag constant, $B$, and the $z$ parameter are determined by the 
bare proton mass, after allowing for the electromagnetic 
self-energy correction $+$0.63 MeV, with the bag radius, $R_p = 0.8$ fm, 
in free space.  
For the neutron, the procedure is the same as that for the proton, 
allowing for the electromagnetic self-energy 
correction, $-$0.13 MeV, but using the values of $B$ and $z$ 
determined above and calculating the d current quark mass and the bag 
radius for the neutron.  
Thus, the u current quark mass ($m_u = 5$ MeV) is the basic input parameter 
used to fix the model parameters so as to reproduce the bare proton and 
neutron masses in free space.  We found $m_d = 9.2424$ MeV in the  
present calculation.  

The coupling constants, $g^q_\sigma$ and $g^q_\omega$, are 
determined so as to fit the saturation properties of 
symmetric nuclear matter \cite{tsu2}.   
In Table \ref{parameters}, SU(2) stands for the parameters and inputs 
obtained and used for the calculation when SU(2) symmetry   
is assumed, namely $m_u = m_d = 5$ MeV.  
We then found: ($g^q_\sigma$, $g^q_\omega) =$ (5.698, 2.744) for CSB,  
and (5.685, 2.721) for SU(2).  For the quark-$\rho$ meson coupling 
constant, to make a realistic estimate, we here use the 
phenomenological value, $g^q_\rho = 4.595$, the value 
at zero three-momentum transfer corresponding to Hartree approximation, 
from Table 4.1 of Ref. \cite{mac}.  (Note that because the QMC model 
does not contain the $\rho$-nucleon tensor coupling \cite{gui}, this gives 
an unrealistically large value for the coupling constant \cite{tsu2}.)  

\subsection*{Proton and Neutron Masses in Nuclear Matter}

As in Equation (\ref{nuclm}), the proton and neutron masses are again 
given by functions of $\sigma$ in matter, and may be expanded in 
terms of $\sigma$ at low $\rho_B$ 
\bg
M_p^{\star} &=& M_p + (3g_{\sigma}^q) \frac{1}{3} 
[2S_{u/p}(0) + S_{d/p}(0)] \sigma + {\cal O}({\sigma}^2),  \label{pexp} \\
M_n^{\star} &=& M_n + (3g_{\sigma}^q) \frac{1}{3}
[S_{u/n}(0) + 2S_{d/n}(0)] \sigma + {\cal O}({\sigma}^2).   \label{nexp} 
\en
Because $m_u \neq m_d$, the u-quark scalar charge is no longer the same as 
the d-quark scalar charge.  We have therefore introduced four kinds of 
quark scalar charges in the expansion: 
$S_{i/j}(\sigma) = \int_{V_j} d{\vec r} \ {\overline \psi}_{i/j} \psi_{i/j}$,  
where $i$ denotes u or d quark, $V_j$ is the volume of $j$ (= p or n)   
and $\psi_{i/j}$ is the $i$ quark wave function in $j$.  
Since the proton consists of 
two u quarks and one d quark, the derivative of $M_p^{\star}$ with respect to 
$\sigma$ is given by $2S_{u/p} + S_{d/p}$.  Similarly, 
the derivative for the neutron is given by $S_{u/n} + 2S_{d/n}$.  

\begin{figure}[b!] 
\centerline{\epsfig{file=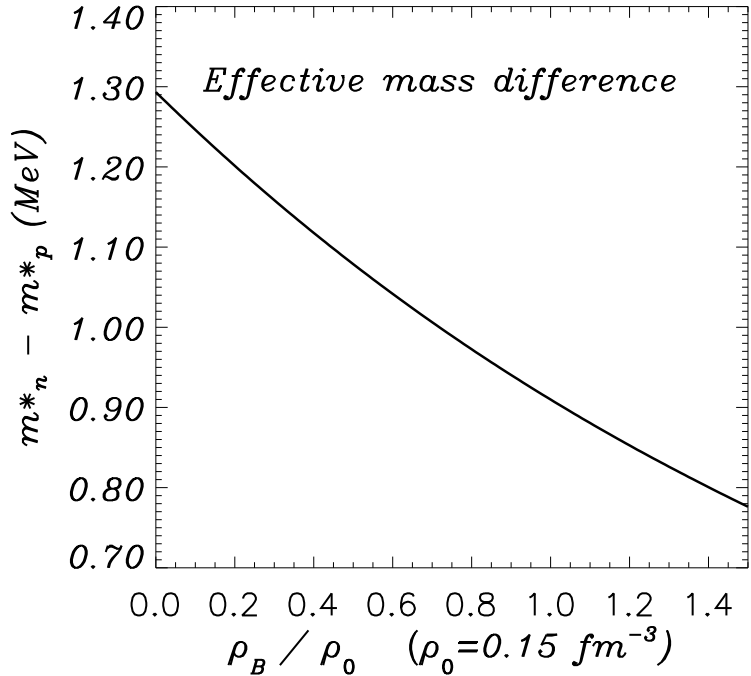,height=10cm}}
\vspace{10pt}
\caption{Neutron-proton effective mass difference in symmetric nuclear matter
with the electromagnetic self-energy corrections.}
\label{fig:onsmass}
\end{figure} 

Taking the difference between the in-medium neutron and proton masses, 
we find 
\bge
\Delta_{np}^{\star} \equiv M_n^{\star} - M_p^{\star} 
= \Delta_{np} - (3g_{\sigma}^q) [S_n(0) - S_p(0)] \sigma + 
{\cal O}({\sigma}^2),  \label{mdif} 
\ene
where $\Delta_{np} = M_n - M_p$, $S_n(0) = \frac{1}{3}
[S_{u/n}(0) + 2S_{d/n}(0)]$ and $S_p(0) = \frac{1}{3}
[2S_{u/p}(0) + S_{d/p}(0)]$.  Here we may expect that 
$S_{u/j} < S_{d/j}$ because the u quark is {\em more relativistic} than 
the d quark in nuclear matter ($m_u < m_d$) --- note that 
the quark scalar charge is given in terms of 
${\overline \psi}_q \psi_q$ in matter.  
Thus, we find that $S_n(0) > S_p(0)$ and then $\Delta_{np}^{\star} 
< \Delta_{np}$ in nuclear medium.  

In Figure \ref{fig:onsmass} we show the neutron-proton effective mass
difference calculated in symmetric nuclear matter, including 
the electromagnetic self-energy corrections.  
One notices that the mass difference 
becomes smaller as the density increases.  
This behavior works in the right direction to resolve the ONS anomaly.   

\subsection*{The ONS Anomaly in Mirror Nuclei}

Now we are in a position to show our results of the
ONS anomaly in mirror nuclei \cite{tsu2}. 
We first present the calculated single-particle energies
for $^{17}$F and $^{17}$O in Table \ref{energylevel}.  
These mirror nuclei
have a common core nucleus, $^{16}$O, and have an extra valence
proton for $^{17}$F and neutron for $^{17}$O.
In order to focus on the strong interaction effect
for the valence proton and neutron,
the Dirac equations are solved without the
Coulomb and $\rho$-meson potentials, or
the electromagnetic self-energy corrections, and keeping
only the charge symmetric $\sigma$ and $\omega$
mean field potentials.  
Consistently, the valence nucleon contributions are not included in the
Coulomb and $\rho$-mean field source densities in the core nucleus. 
However, for the nucleons in the core nucleus, electromagnetic self-energy
corrections and the Coulomb potential as well as the $\rho$ mean field
potential are included in addition
to the $\sigma$ and $\omega$ mean field potentials in solving
the Dirac equations.
Results are shown for two cases in Table \ref{energylevel}: 
calculation with charge symmetry breaking (denoted by CSB) and 
calculation performed assuming SU(2) symmetry (denoted by SU(2)).  

\begin{table}
\caption{Calculated single-particle energies (in MeV) for $^{17}$F 
and $^{17}$O.}
\label{energylevel}
\begin{tabular}{r|cc|cc}
 &CSB & &SU(2) & \\
 &$^{17}$F &$^{17}$O &$^{17}$F &$^{17}$O \\
\tableline 
p 1s$_{1/2}$ &-28.800 &-28.805 &-28.663 &-28.663 \\
  1p$_{3/2}$ &-14.154 &-14.158 &-14.032 &-14.032 \\
  1p$_{1/2}$ &-12.495 &-12.499 &-12.383 &-12.383 \\
\tableline 
n 1s$_{1/2}$ &-33.367 &-33.372 &-32.967 &-32.967 \\
  1p$_{3/2}$ &-18.259 &-18.263 &-17.918 &-17.918 \\
  1p$_{1/2}$ &-16.587 &-16.590 &-16.258 &-16.258 \\
\tableline 
valence &p         &n        &p        &n        \\
1d$_{5/2}$ &-3.918 &-4.099  &-3.848  &-3.848  \\
\end{tabular}
\end{table}

The SU(2) results for $^{17}$F and $^{17}$O agree perfectly  
with each other as they should.   
Single-particle energies in the cores of $^{17}$F and 
$^{17}$O are slightly different for CSB. 
This difference is induced by the different (effective) masses for 
the valence proton and neutron, arising from the charge and density 
dependence of their coupling to the self-consistent scalar mean field. 
This also causes a second order effect on the Coulomb and $\rho$-meson 
potentials through the self-consistency procedure. 

It is interesting to compare the binding energy differences  
between the valence proton in $^{17}$F and neutron in $^{17}$O. 
In CSB, the result gives,  
$E(p)(1d_{5/2}) - E(n)(1d_{5/2}) \simeq 0.18$ MeV, while the SU(2) case 
is zero as it should be. 
This amount already shows a magnitude similar to that of the 
observed binding energy differences.  

\begin{table}
\caption{Calculated single-particle energies of mirror nuclei. 
For each nucleus, the top row shows the single-particle energy of the  
valence proton or neutron (the orbit is also indicated). 
$\delta E_\rho$ stands for the contribution from   
the $\rho$-meson central and spin-orbit potentials of the core nucleus.  
The discrepancies between the experimental 
values and the theoretical expectations in the absence of 
charge symmetry violating strong interactions 
are taken from Table II of Ref. \protect\cite{sha}, by averaging 
over the theoretical values.}
\label{summary}
\begin{tabular}{c|cc|cc}
 &$^{15}$O(p)&$^{15}$N(n)  &$^{17}$F(p)&$^{17}$O(n) \\
\tableline
1p$_{1/2}$ or 1d$_{5/2}$(MeV)  &-14.397 &-14.631  &-3.918  &-4.099 \\
$\delta E_\rho$(MeV)        &-0.055  &0.056  &-0.005  &0.005 \\
Total(MeV)           &-14.452 &-14.575  &-3.923  &-4.094 \\
\tableline
$\delta E=E(p)-E(n)$
 &$\delta E =$&123(keV)
 &$\delta E =$&171(keV) \\
 &observed =&230(keV) &  observed =&220(keV)     \\
\tableline
 &$^{39}$Ca(p)&$^{39}$K(n) &$^{41}$Sc(p)&$^{41}$Ca(n) \\
\tableline
1d$_{3/2}$ or 1f$_{7/2}$(MeV) &-16.407 &-16.689 &-6.970  &-7.210  \\
$\delta E_\rho$(MeV)        &-0.087  &0.088   &-0.006  &0.006  \\
Total(MeV)           &-16.494 &-16.601   &-6.976  &-7.204 \\
\tableline
$\delta E=E(p)-E(n)$
 &$\delta E =$&108(keV)  &$\delta E =$&228(keV) \\
 &observed =&340(keV)    &observed =&460(keV)     \\
\end{tabular}
\end{table} 

In Table \ref{summary}, we summarize the calculated single-particle energies
for the valence proton and neutron of several mirror nuclei 
(in CSB) \cite{tsu2}.  
Comparing the $\rho$-potential contributions for the hole states with  
core plus valence states, one notices the shell effects due to 
the $\rho$-potentials.  
These results reflect the difference in the shell structure, 
namely the hole states tend to have larger $\rho$-potential 
contributions than the core plus valence nucleon states. 

The binding energy differences obtained indicate that the prime  
CSB effects originate in the u-d current quark 
mass difference. 
The calculated binding energy differences give 
of the order of about a few hundred keV. This is precisely the  
order of magnitude which is observed as 
the ONS anomaly \cite{mil,sha}.  

\section*{SUMMARY}

Using the QMC model, we have discussed CSB in 
nuclear medium and calculated the ONS anomaly in mirror 
nuclei, including the quark degrees of freedom explicitly.  
We stress that the present contribution to the ONS anomaly 
is based on a very simple but novel idea, namely the slight difference 
between the quark scalar densities of the u and d quarks in a bound nucleon, 
which stems from the u and d quark mass difference \cite{sai,tsu2}. 
This implies that the in-medium proton-$\sigma$ and 
neutron-$\sigma$ coupling constants differ from their 
values in free space and that the neutron-proton effective 
mass difference is reduced in matter.  

Our results were obtained within an 
explicit shell model calculation, based on quark degrees of 
freedom. They show that once CSB 
is set through the u and d current quark mass difference so as to reproduce  
the proton and neutron masses in free space, it can produces 
binding energy differences for the valence 
(excess) proton and neutron of mirror nuclei of a few hundred keV. 
The origin of this effect is 
so simple that it is natural to conclude that a sizable fraction 
of CSB in mirror nuclei arises from 
the density dependence of the u and d quark scalar densities in a bound 
nucleon.  

It is a fascinating challenge for the future to compare this result with the 
traditional mechanism involving $\rho-\omega$ mixing \cite{blu}. This 
will involve the issue of the possible momentum dependence of the 
$\rho-\omega$ mixing amplitude \cite{mil,gol}.  In addition, one would 
need to examine whether there is any deeper connection between these 
apparently quite different sources of charge symmetry violation. 

\section*{ACKNOWLEDGMENTS}

The author would like to thank K. Tsushima, A.W. Thomas and A.G. Williams  
for valuable discussions.  
This work was supported by the Australian Research Council and the Japan
Society for the Promotion of Science.  In the present paper, 
Fugure 1 was reprinted from Nucl. Phys. A 609 (1996), Saito et al., 
``Self-consistent description of finite nuclei based on a relativistic 
quark model'', p.352 (Figure 4), and Figure 2 and Table 1-3 were reprinted 
from Phys. Lett. B 465 (1999), Tsushima et al., ``Charge symmetry 
breaking in mirror nuclei from quarks'', p.38 (Figure 1 and Table 1), p.39 
(Table 2) and p.41 (Table3), with permission from Elsevier Science.

\end{document}